\documentclass{aa}            
\usepackage{graphicx}
\begin{document}
\thesaurus{ }     
\title{Near Infra-red Spectroscopy of V838 Monocerotis}
\author{D. P. K. Banerjee and N. M. Ashok}
\offprints{D. P. K. Banerjee}
\institute{Physical Research Laboratory, Navrangpura, \\
	      Ahmedabad -- 380 009, India\\
              email: orion, ashok@prl.ernet.in}
\date{Received 10 June 2002 / Accepted 27 August 2002}
\titlerunning{IR spectroscopy of V838 Mon}
\maketitle

\begin{abstract}
Near - IR, multi - epoch, spectroscopic and photometric observations of the 
enigmatic, eruptive variable V838 Mon in the $JHK$ bands are reported. 
One of unusual features of the spectra is the detection 
of several strong  neutral TiI lines in emission in the $K$ band. From
the strength of these lines, the mass of the ejected envelope is
estimated to be in the  range 10$^{-7}$ to 10$^{-5}$ $M$$_\odot$.
 The spectra also show the strong presence of the first and second 
overtone $^{12}$CO bands seen in the $K$ and $H$ bands. The CO bands show 
a complex evolution. Deep
water bands at 1.4 ${\rm{\mu}}$m and 1.9 ${\rm{\mu}}$m are also seen later 
in the object's evolution. Blackbody fits to the $JHK$ photometric data
 show that V838 Mon has evolved to   temperatures between 2400-2600 K
 by $\sim$ 130 days after outburst. The spectra at this stage have the
general characteristics of a very cool M giant.

      \keywords{Stars: individual: V838 Mon - Stars: supergiants - Infrared:
       stars - Stars: novae - Techniques:	 spectroscopic}

   \end{abstract}

\section{Introduction}
The eruptive variable V838 Mon was first reported to be in outburst
 on 6 January 2002 by  Brown (2002). A first maximum ($V$ $\sim$ 9.8) in its
 lightcurve  was reached around 11 January. Subsequently there was a slow 
 steady decline followed by  a second strong outburst on 2 February which
 changed the brightness by  4.3 magnitudes  to a peak value of 
 $V$ $=$ 6.7. The overall evolution   of the light 
curve has thus been complex. The initial spectra of the object in the 
optical have shown several emission lines, in general having P Cygni 
profiles, of BaII, LiI and those of several $s$ process 
elements (Munari et al. 2002a). However, the most striking development was
 the  detection of an expanding light echo around the object seen most
 clearly in $U$ band images
(Munari et al. 2002a). Being a rare phenomenon,
this heightened interest in an already puzzling object, resulting in 
several IAU circulars on the subject and also warranting 
special Director's Discretionary Time allocation  on the HST
(Bond et al. 2002). The light echo seems to be caused by scattered light
from dust shells which existed even before the outburst. This is supported 
by the fact that the progenitor of V838 Mon was detected by IRAS in the 60
and 100 ${\rm{\mu}}$m bands indicating the presence of low temperature dust.
 In this work, we  present   results from 
$JHK$  observations of V838 Mon which have been made at five, fairly
 evenly-spaced epochs and which should help in following the temporal
  evolution and understanding the nature  of V838 Mon  -  questions
of considerable   importance and interest at present. 

\section{Observations}

 Near-IR $JHK$ spectra at a resolution of $\sim$ 1000 were obtained at the
 Mt. Abu 1.2m  telescope using a Near Infrared Imager/Spectrometer
 with a 256$\times$256 HgCdTe NICMOS3 array. We present here the spectroscopic
 observations  of five days viz. 2 February, 25 March, 9 April, 
 2 May and 14 May 2002. It may be noted that the 2 February spectra were
 acquired between 2.729 - 2.762 UT  and coincide with the time when the
  second outburst of V838 Mon was  just underway as indicated by VSNET
  reports. VSNET reports, centered around the second outburst, give photometric
  values  of $V$ = 10.67, 10.708, 8.193 and 8.02 at epochs of February 
  1.487, 1.858,  2.799 and 2.913 UT from which an idea can be
   obtained when the  outburst began.  
 In each of the $J,H$ and $K$ bands a set of at least two spectra were taken 
 and sometimes as many as ten. In each set the  star was offset to two 
 different position of the  slit (slit width $=$ 2 arc sec.\ ). The 
 signal to noise ratio of the spectra, as determined using IRAF, is moderate
  and ranges between 30 - 50 in $J$, 30 - 60 in $H$ and 30 - 80 in the $K$ band.
 The exposure times for the spectra presented here are as follows
 (given in order of $J,H$ and $K$): 2 Feb - 120, 90, 90s; 25 March
 - 15, 10, 15s; 9 April - 15, 10, 15s; 2 May - 120, 45, 45s and 14 May 
 - 120, 60, 60s. Spectral calibration   was done using the OH sky lines 
 that register with the spectra.  The comparison stars that were used for
 ratioing the spectra, were either HR 2714 or HR 3314 in all cases. The spectra
 of the comparison stars were taken at similar air-mass as that of V838 Mon
 and the ratioing process thereby removes the telluric lines.
 
\begin{figure}[t]
\centering
\includegraphics[bb=23 118 513 617,width=3.2in,height=4.5in,clip]{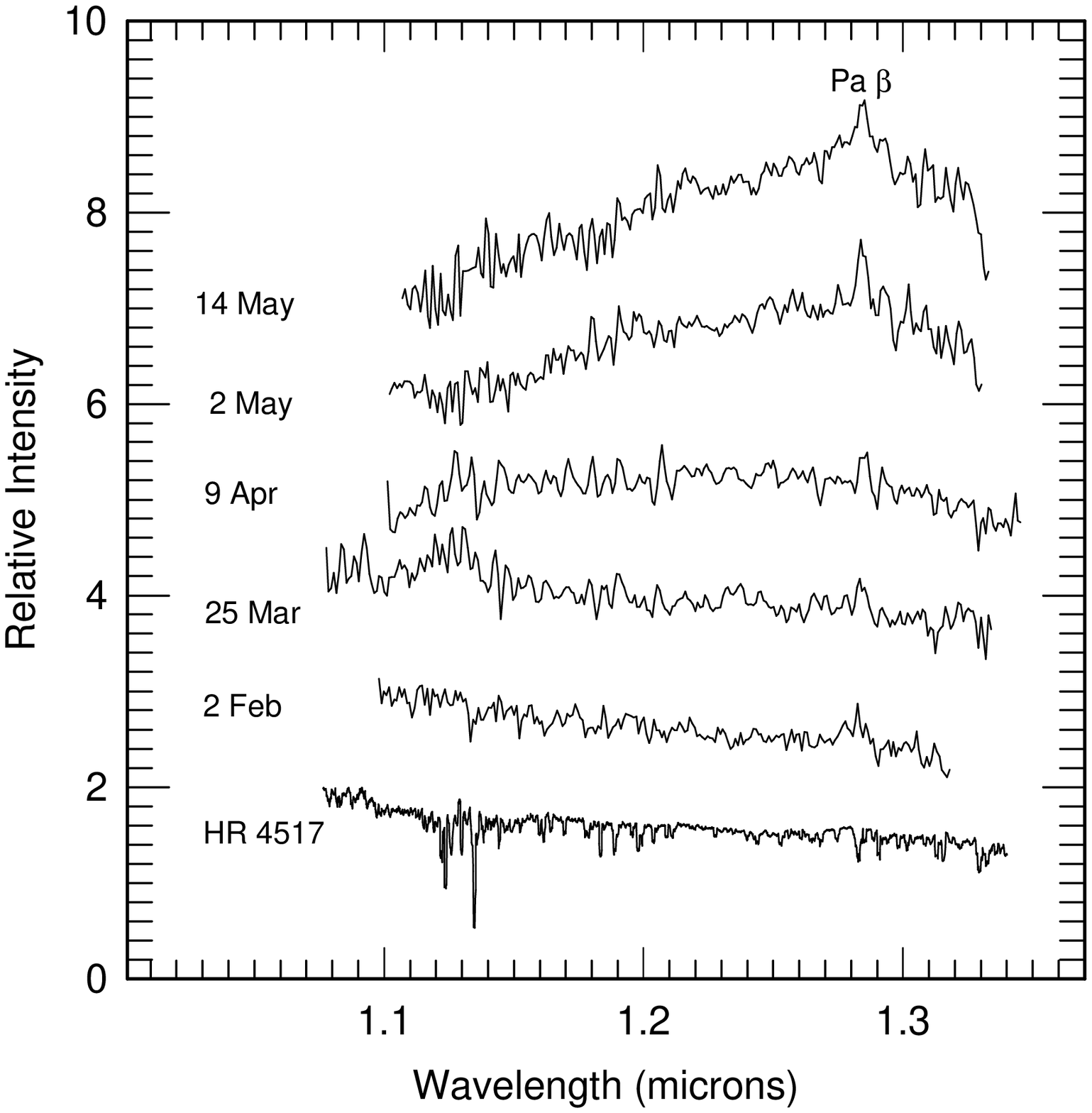}
\caption[]{ The $J$ band spectra of V838 Mon are shown at different epochs. 
The spectra have been offset from each other for clarity. The comparison
spectra  of HR4517 is from Wallace et al.(2000).}
\label{fig1}
\end{figure}
 
 Photometry in the $JHK$ bands was  done on 3 and 14 May using the NICMOS3 
 array, mentioned above, in the imaging mode. A large number of dithered frames
  in 4 positions offset by 30 arcsec were obtained in all the filters. The
  times for all the individual frames ranged between 70 and 250ms. 
 The sky frames were generated using these dithered frames. The mean
air-mass at the time of observations for V838 Mon was 1.79 on 3 May and 
2.04 on 14 May. HR 2714 ($\delta$ Mon, $V$= 4.15, spectral type A2V) was used as
the standard star on both the days and observed soon after V838 Mon at
similar airmasses. The atmospheric extinction corrections were done using 
average values of $k$$_{\rm J}$ = 0.15, $k$$_{\rm H}$ = 0.15 and
$k$$_{\rm K}$ = 0.1 mag for the Mt. Abu observatory site. The adopted $JHK$
magnitudes for HR 2714 ($J$ = 4.06, $H$ = 4.03, $K$ = 4.02) were
calculated from the colors of an A2V star as given by Koornneef (1983).  The
near-IR  photometric  and spectroscopic data were all reduced using IRAF.

\begin{figure}[t]
\centering
\includegraphics[bb=52 129 530 625,width=3.2in,height=4.5in,clip]{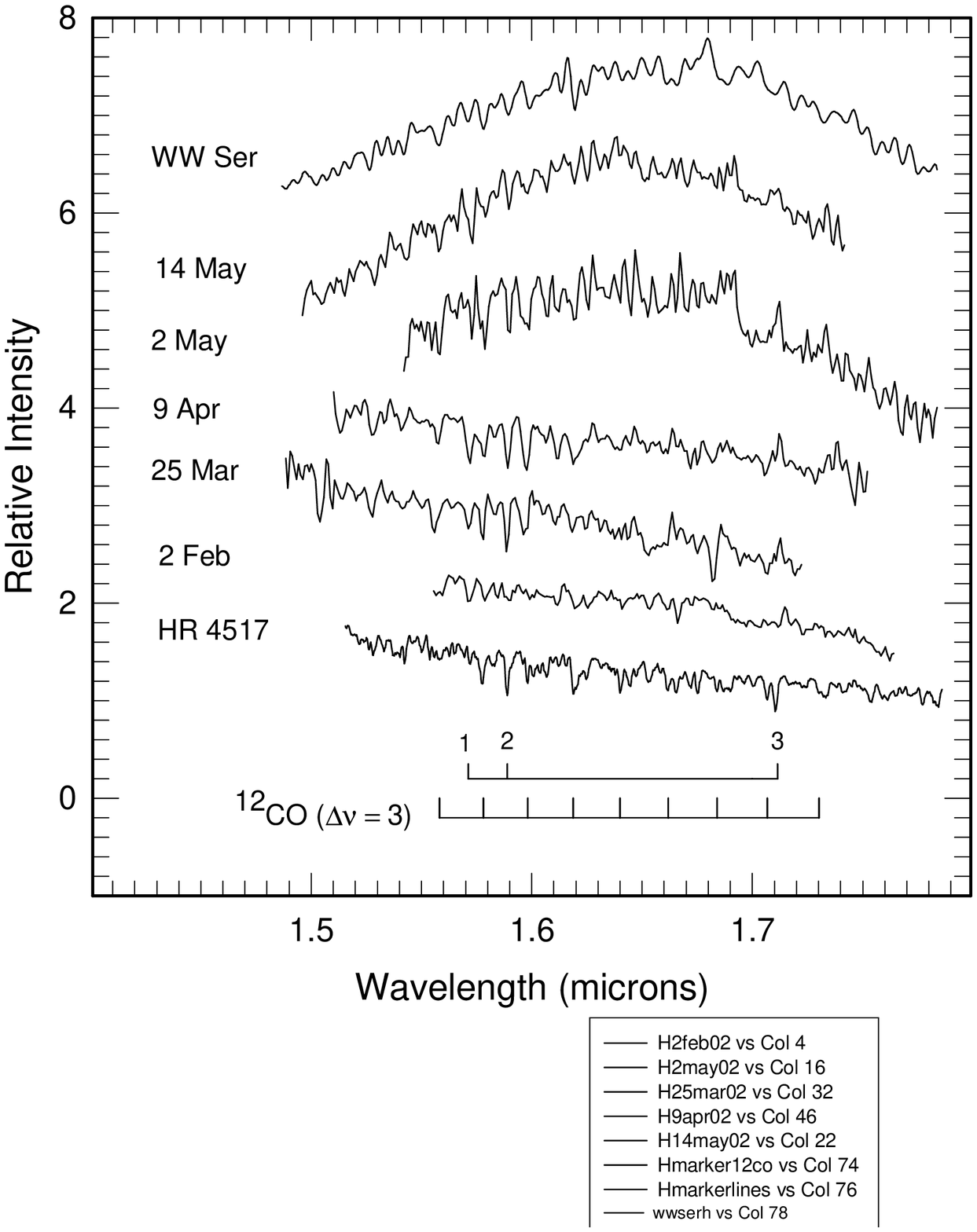}
\caption[]{ The $H$ band spectra of V838 Mon are shown at different epochs. 
The spectra have been offset from each other for clarity. The comparison
spectra  of HR4517 is from Meyer et al. (1998) while that of WW Ser is from 
Lancon \& Roca-Volmerange (1992). The marked features are discussed in the
 text.}
\label{fig2}
\end{figure}


\begin{figure}
\centering
\includegraphics[bb=35 160 515 659,width=3.2in,height=4.5in,clip]{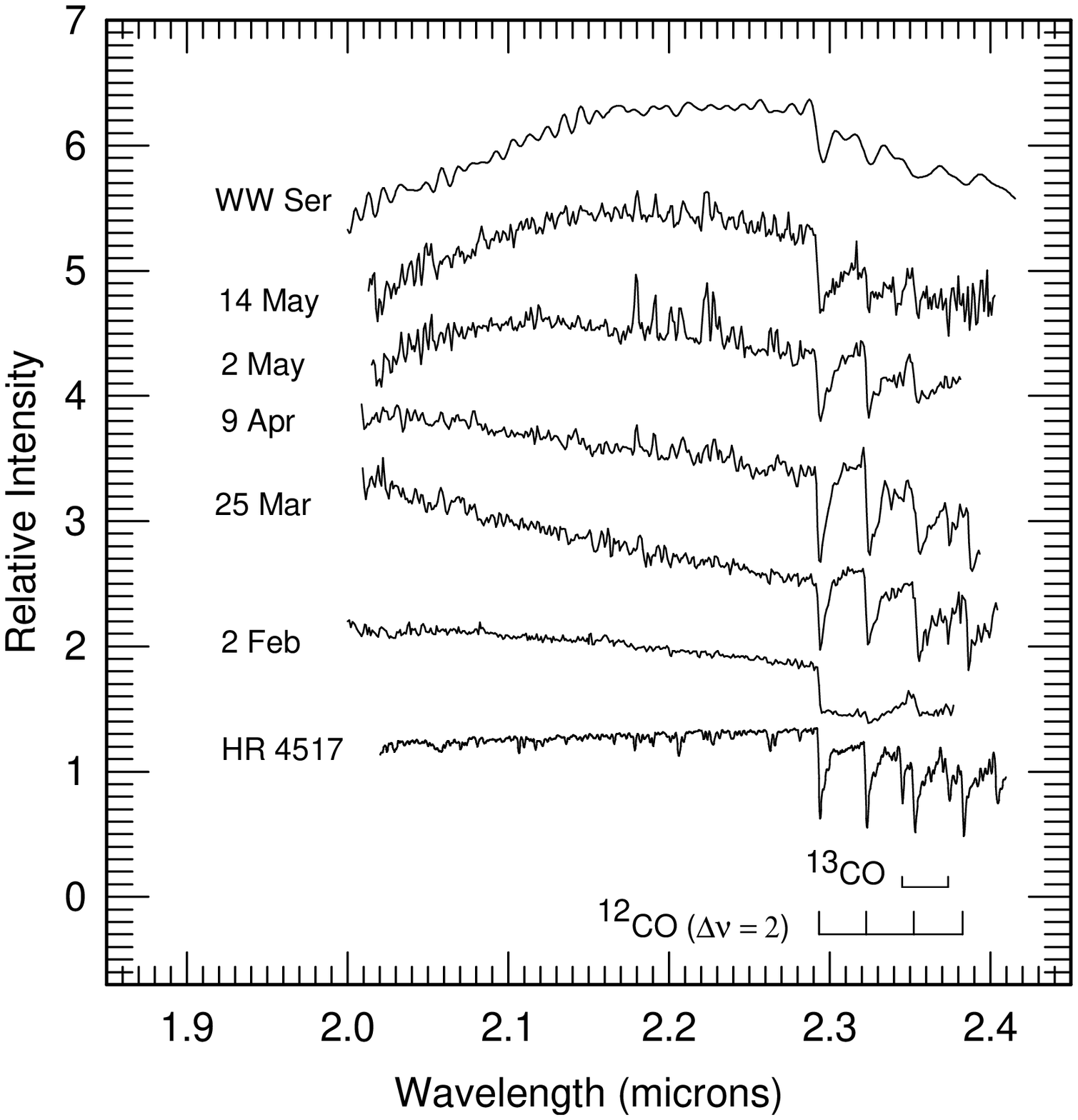}
\caption[]{The $K$ band spectra of V838 Mon are shown at different epochs. 
The spectra have been offset from each other for clarity. The comparison
spectra ( see text) of HR 4517 is from Wallace \& Hinkle (1997) while
 that of WW Ser is from 
Lancon \& Roca-Volmerange (1992). The $^{12}$CO bands and also the TiI lines 
around 2.2 ${\rm{\mu}}$m can be seen.}
\label{fig3}
\end{figure}

\section{Results}
\subsection{Near-IR $JHK$ band spectra}
The $JHK$ spectra are shown in Figs. 1, 2 and 3.  We first discuss the spectral
 features that are observed and later discuss the shape of the
 continuum in the $JHK$  bands. We have compared our spectra, with the aim of 
 classification and line identification, with the spectral classification
  catalogs of $JHK$  spectra  compiled  by Meyer et al. (1998),
 Wallace \& Hinkle (1997), Lancon \& Roca-Volmerange (1992), Wallace et al.
 (2000) 
 and Forster Schreiber (2000). The spectra from some of these last-mentioned
references are available in electronic form from the ADC \& CDS data centers
and also from the site http://www.noao.edu/archives.html.
  To guide the reader in interpretation/comparison of our observed data,
  we have added  two reference spectra  to Figs. 1, 2 and 3 
 - one at the bottom and the other at the top. The  selected comparison
  spectra of HR 4517 (spectral type M1 III) and WW Ser (M8 III) are close to
  the spectral  type of V838 Mon as on 2 February and 14 May.  However it
   was not possible to add a comparison $J$ band spectra for 14 May 
  because V838 Mon had cooled to  spectral  class  M8 or 9 by then and similar
 comparison spectra were not available. It must be pointed out that the 
 spectra of HR 4517  in Figs. 1, 2 and 3 have been
 smoothed by a seven point moving average to degrade them somewhat from a higher
 resolution (R = 3000) to a comparable resolution as our observations. 
 Further the slope of the $J$ and $H$ band spectra of HR 4517 have been
  tilted to match those of 
 our spectra because in the original data the spectra had been normalized to be
 flat.  The slope correction does not alter the spectral features at all and
 the moving-average smoothing ( instead of a convolution with the instrument
 function) should adequately serve the primary purpose of enabling a comparison
 between spectral features. The spectra for WW Ser, which are from Lancon \&
 Roca-Volmerange (1992) are at a resolution of 500.\\ 
 
In the $J$ Band, the only feature which appeared weakly but persistently
in emission is Paschen beta at 1.2818 ${\rm{\mu}}$m. We give the 
equivalent  widths for this line in case they can be used in
conjunction with  other Hydrogen line data (obtained in other studies) to
study some properties of V838 Mon ( for e.g. interstellar extinction 
towards it). The measured equivalent widths are 2.7, 5.0, 5.9, 8.3 and 6.9$\AA$
on 2 February, 25 March , 9 April, 2 May and 14 May respectively. The
errors  in measurement typically lie between $\pm$10 to 20 percent of the 
measured values.
   The $H$ band spectra show a multitude of spectral features  which are rather
  characteristic of the spectrum of  cool  M type giant stars. Some of the
 principal spectral features listed by Meyer et al. (1998) which characterize
the $H$ band spectra of stars are also seen here. Among these are the
second overtone $^{12}$CO ($\Delta$$v$ $=$ 3) bands whose positions 
 are marked in Fig. 2. Although weak, the $v$ = 3-0 (1.558 ${\rm{\mu}}$m),
 4-1 (1.578 ${\rm{\mu}}$m),  5-2 (1.598 ${\rm{\mu}}$m) and
  6-3 (1.619 ${\rm{\mu}}$m)  bands are  consistenly seen. The positions of the
 higher vibrational transitions  are also marked in the diagram but they are
 not clearly present in the data.  Three other  features (marked 1, 2, and 
 3 in Fig. 2)  are also persistent. Feature 1 at 1.5711 ${\rm{\mu}}$m could 
 be the HI Brackett 15 line while  feature 2  at 1.5892 ${\rm{\mu}}$m is 
 attributable to SiI (Forster Schreiber 2000).  Feature 3 is seen in
 emission at all phases and although it coincides well with the
 position of MgI 1.7113 ${\rm{\mu}}$m it is expected to be in absorption had it 
 been due to MgI (Meyer et al. 1998). The spectra in the $H$ band, 
 in general, resemble those of later M type giant stars, but not completely.
 However it must be pointed out that the spectra of V838 Mon is a superposition
 of photospheric absorption lines and also emission lines from the ejecta and 
 their coaddition  can cause deviations from standard stellar spectra.

\subsection{TiI lines and the mass of the envelope}
 The $K$ band spectra have two striking features. The first of these are 
 several emission lines seen around 2.2 ${\rm{\mu}}$m which were first visible
 on 9 April, peaked in strength on 2 May and again weakened by 14 May
 . A magnified section around these lines is shown in Fig. 4. The  
preliminary  indication  that they could be  due to TiI  was their presence  
 in absorption, at similar wavelengths, in  the high resolution  spectra 
 of Arcturus ( Hinkle et al. 1995). We believe
  these lines are  due to neutral Titanium (Banerjee \& Ashok 2002) after
  having matched the observed wavelengths of these lines with the
 laboratory spectrum of TiI  as given by Forsberg (1991). 
 It is quite remarkable that all the
 significantly strong lines listed by Forsberg (1991) in the $K$ band appear 
in our spectrum. The laboratory wavelengths of these lines are marked
in Fig. 4. As can be seen there is a good match. The intensities of 
the lines also match fairly well the intensities given by Forsberg (1991).
We therefore feel the identification is secure. The detection of TiI 
emission lines in the $K$ band is certainly rare in eruptive variables
 or even in 
astronomical objects - in fact it's unclear  whether these lines have 
ever been seen before and their presence adds to the mystery of V838 Mon.
 Being in emission, they cannot be of photospheric
origin (not in a cool object) and must therefore arise from the circumstellar
ejecta. Except for the 2.2627 ${\rm{\mu}}$m line which arises from a transition
 between the $a$$^{3}$G$_{4}$ - $z$$^{3}$F$_{4}$  levels of the
   3$d$${^{3}}$($^{2}$G)4$s$ - 3$d$${^{2}}$($^{3}$F)4$s$4$p$($^{3}$P)
configuration, all the other lines arise from transitions between different
 levels (not listed here) of the  
3$d$${^{3}}$($^{4}$P)4$s$ - 3$d$${^{2}}$($^{3}$F)4$s$4$p$($^{3}$P)
   configuration. The excitation energies for the lower levels 
 of all the TiI lines shown in Fig. 4 range between 1.734 to 1.749 ev 
 except for the 2.2639 ${\rm{\mu}}$m for which it is 
 1.879 ev (the ionization potential of TiI is 6.83 ev). Thus all these lines
  emanate from low-lying energy levels vis-a-vis the ground state and
  suggest that they may be collisionally excited in a low temperature gas.
  The heating of the gas is likely to be caused by the mass outflow from 
  V838 Mon during the  present outburst impinging on pre-existing  matter.
  The indication for such pre-existing matter is suggested from the
   light-echo and the evidence for a mass outflow from V838 Mon is confirmed 
   by the strong P-Cygni profiles seen in its optical spectra. The above 
   mechanism of excitation can also be the
 cause for the NaI doublet at 2.2056 ${\rm{\mu}}$m and 2.2084 ${\rm{\mu}}$m,
 which appear blended into a broad feature (marked in Fig. 4)
  and for which the lower level excitation energy is  3.19 ev.
 From Forsberg's (1991) line list, 3 other lines of TiI are also
  expected to be seen  at 2.2896, 2.2970 and 2.3348 ${\rm{\mu}}$m.
 However the position of these lines coincide with the locations of the
 deep $^{12}$CO
 bands observed and hence their presence can be masked.
  There are
  certain other features which remain unidentified   in the $K$ band.
 Among these is a cluster of five closely spaced emission lines at 2.0325,
  2.0374,  2.0427, 2.0483 and 2.0515 ${\rm{\mu}}$m which 
 are weak but clearly seen in both the 2 and 14 May spectra. It may be
  pointed out that at these line positions there are no matching  lines 
 (unlike the TiI lines)  in the Arcturus spectral atlas (Hinkle et al., 1995).

\begin{figure}[t]
\centering
\includegraphics[bb=70 230 343 599,width=2.75in,clip]{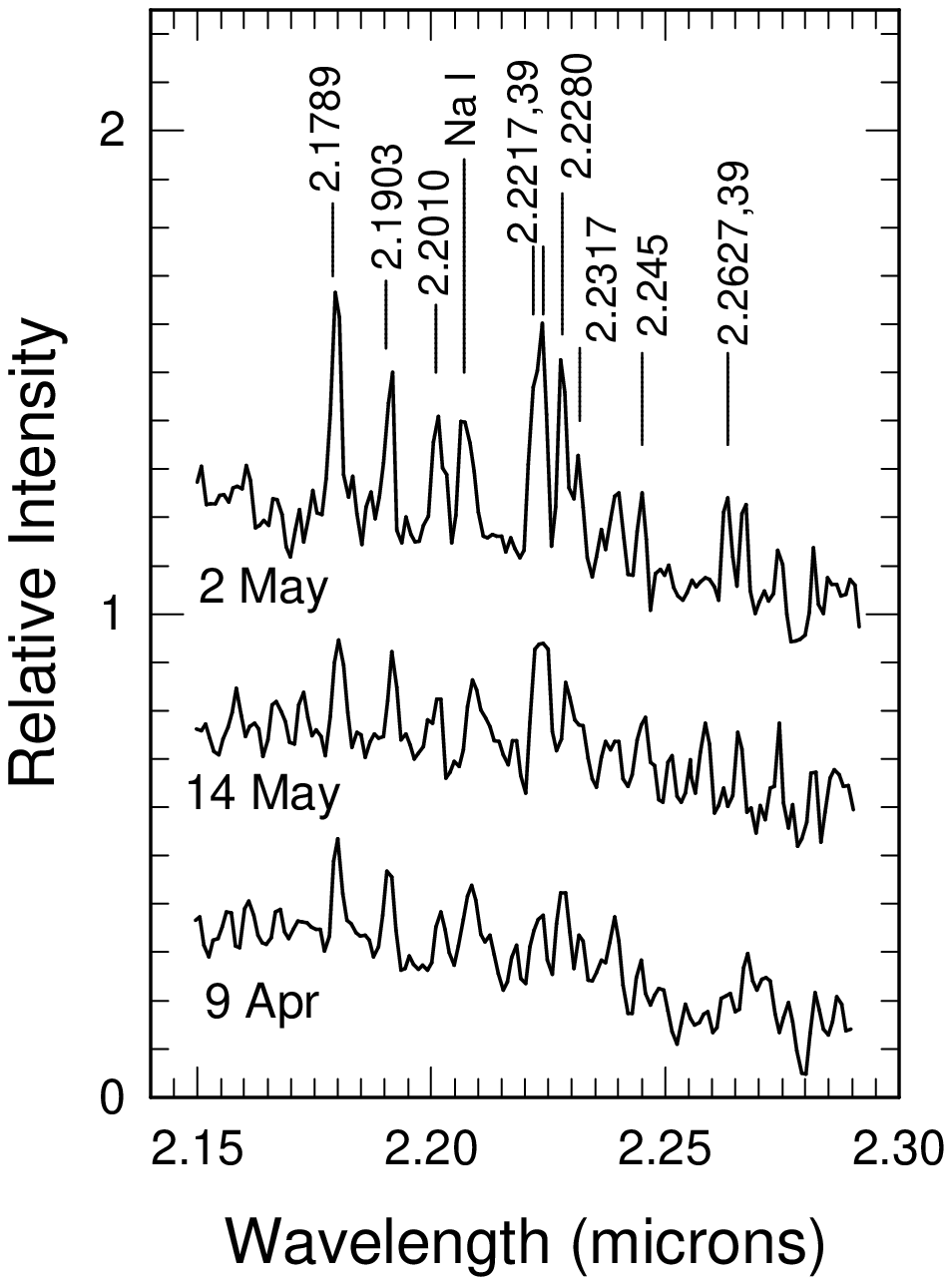}
\caption[]{ A magnified section of the $K$ band spectra showing the
 unusual presence of several TiI lines in V838 Mon.}
\label{fig4}
\end{figure}

 From the observed strength of the  TiI lines we estimate the amount of
  TiI in the ejecta and from this calculate the mass of the shell. The
 observed luminosity in an individual TiI line is first calculated which
  will be given by:
 
 \begin{equation}
 L{_{\rm line}} = 4{\pi d{^{\rm 2}}}FW {\rm ~ergs/s}   
 \end{equation}
 \vskip 2mm
 where $W$ is the observed equivalent width  of the line (in Angstroms)
  and $d$ is the  distance to V838 Mon. $F$ is the observed continuum 
 flux in ergs/cm$^2$/s/$\AA$   at the line center. Since the
  TiI lines are situated  close to the $K$ band center (2.2 microns), the 
 observed $K$ band magnitude of V838 Mon has been used for all the lines to
 calculate the flux $F$. Photometric results of the $K$ band magnitudes
  which are used here are presented  later. 
 
 If the line originates in a downward transition from level 2 to 1 then,
 $L$$_{\rm line}$ is also given by:
 
 \begin{equation}
 L{_{\rm line}} = (N{_{\rm 2}}/N(Ti))N(Ti)A{_{\rm 21}}h \nu V {\rm ~ergs/s} 
  \end{equation}
    \vskip 2mm
  where $N$$_{\rm 2}$ is the number density  of TiI atoms in the
 excited state  2, $N(Ti)$ is the total number density of the TiI atoms,  
  $A$$_{\rm 21}$ is the Einstein  coefficient of spontaneous emission,
  $h$$\nu$ is the energy of the emitted photon  and $V$ is the volume of the
  emitting gas.

  By equating Eq. (2) to Eq. (1) the number of TiI atoms in the shell
  ( equal to $N(Ti)V$) 
  can be found.The two unknown quantities of Eq. (2) viz. $A$$_{\rm 21}$
and   $N$$_{\rm 2}$/$N(Ti)$  are found  as follows.  
   The value of $A$$_{\rm 21}$  for any particular line is given by
    
  \begin{equation}  
  A{_{\rm 21}} =
0.667 \times 10{^{\rm 8}}gf/(g{_{\rm 2}}\lambda{^{\rm 2}}) {\rm ~s}{^{\rm -1}}
 \end{equation}
 \vskip 2mm
 \begin{table*}[t]
\caption[]{The table gives the number of TiI atoms ($N(Ti)V$ in units of
10$^{44}$) calculated for the different lines. The equivalent widths and
 $N(Ti)V$ values marked with superscripts 1 and 2 refer to a blend of
lines (see text). E$_1$ and E$_2$ are the energy values for the lower and upper
levels of the transitions.}
\begin{tabular}{llllllllllllllllllllllllll}
\hline \\ 
$\lambda$(${\rm{\mu}}$m)  &&&&  $W$(\AA)    &&&& log($gf$) &&&& $g$$_2$ &&&&
 {E$_1$}(cm$^{-1}$) &&&& {E$_2$}(cm$^{-1}$) &&&&     $N(Ti)V$ \\
\hline 
\hline \\ 
2.1789  &&&& 8.06   &&&& -1.161  &&&&  9 &&&& 14105.68 &&&& 18695.23 &&&& 0.43 & \\
2.1903  &&&& 5.88   &&&& -1.449  &&&&  7 &&&& 14028.47 &&&& 18593.99 &&&& 0.59 & \\
2.2010  &&&& 4.86   &&&& -1.877  &&&&  5 &&&& 13981.75 &&&& 18525.07 &&&& 1.30 & \\
2.2217  &&&&        &&&& -1.77   &&&&  3 &&&& 13981.75 &&&& 18482.86 &&&&  & \\
2.2239  &&&& 23.8$^1$  &&&& -1.658  &&&&  5 &&&& 14028.47 &&&& 18525.07 &&&& 1.36$^1$  & \\
2.2280  &&&&        &&&& -1.756  &&&&  7 &&&& 14105.68 &&&& 18593.99 &&&&  & \\
2.2317  &&&&        &&&& -2.124  &&&&  1 &&&& 13981.75 &&&& 18462.83 &&&&  & \\
2.245   &&&& 2.83   &&&& -2.251  &&&&  3 &&&& 14028.47 &&&& 18482.86 &&&& 1.87 & \\
2.2627  &&&& 2.02$^2$   &&&& -2.607  &&&&  5 &&&& 14105.68 &&&& 18525.07 &&&& 2.29$^2$ & \\
2.2639  &&&&        &&&& -2.85   &&&&  9 &&&& 15156.79 &&&& 19573.97 &&&&  & \\
 
\hline
\end{tabular} 
\end{table*}
  
\noindent where $gf$ is the weighted oscillator strength for the transition,
$g$$_{\rm 2}$ is the statistical weight of  level 2 ( equal to 2$J$ + 1) and
$\lambda$ is the wavelength  in microns. 
The log($gf$) and $g$$_{\rm 2}$ values for all the TiI lines observed here
 have been listed in a   compilation by Kurucz 
 (http://kurucz.harvard.edu/linelists.html) 
 and these values have been used to determine $A$$_{\rm 21}$. To 
 estimate $N$$_{\rm 2}$/$N(Ti)$ we make the assumption that the gas is in 
 local thermal equilibrium and therefore the population of atoms 
in different excited levels can be described by a Boltzmann distribution i.e.
   
   \begin{equation}
   N{_{\rm 2}}/N(Ti) = (g{_{\rm 2}}/U)e{^{(\rm -\chi{_{02}}/kT)}}
   \end{equation}
   \vskip 2mm
   where $U$ is the partition function, $T$ is the    temperature of the gas 
   and $\chi$${_{\rm 02}}$ is the energy difference  between the ground state
    and level 2. The energy level values for   the different states have
 been taken from  Kurucz's line list. The partition function values, for
  different temperatures, have been taken from Allen (1976) and Aller (1963). 
  
  The number of TiI atoms in the shell, as computed from Eqs. (1) to (4), 
  is listed in Table 1 along with some of the other relevant parameters needed
  in the   calculations. The $W$ values in Table 1 are for the 2 May 2002 
 observations when the TiI lines were strongest and the signal-to-noise ratio 
 in the continuum was about 50.  To reduce errors in the measured 
  equivalent widths, an average
  $W$ was measured from  at least two (at times four) different spectra.
 However, as can be seen from Fig. 4, the 2.2217, 39, 80 and 2.2317 lines are 
 blended ( blend 1 ) and so also the 2.2627 \& 2.2639 lines 
 (blend 2) and so the individual equivalent widths cannot be measured. 
 To calculate $N(Ti)V$ for these lines is still possible by the following
 method. The combined equivalent width of blend 1 and 2  was measured 
 separately. The
 line luminosity corresponding to this  combined equivalent  width for a blend
 ( as calculated from Eq. (1)) will be equal to the summation 
 of the right hand side of Eq. (2) for all the blended lines in the
  group. Thus an average $N(Ti)V$ for the blended lines can be determined.
  The equivalent widths for the blend 1 and  2 lines in Table 1 is therefore
 the combined equivalent width. In the calculations we have adopted a value 
 of $d$ = 790 pc (Munari et al., 2002a), $T$ = 3600 K and $U$  = 23
  ( for T = 3600 K).
  The  temperature of the shell is not  known, but is possibly low, in the range
 3000 to 5000 K. Such a possibility is indicated  because, as mentioned 
 earlier, the TiI lines are all low excitation lines. Secondly the presence 
 of CO which is seen in  emission  in the $K$ band (discussed subsequently)
 is generally associated with this temperature range. 
     
From Table 1 it is seen that $N(Ti)V$, as derived from the different lines, 
is reasonably consistent with a variation of $\sim$ 5 among the 
individual values. We use a mean value of
 $N(Ti)V$ = 1.31$\times$$10^{44}$
for calculating an important parameter viz. the mass of the shell. 
Assuming that the $N(Ti/H)$  and $N(Ti/He)$ abundances in the shell are
adequately represented by cosmic abundances (Allen, 1976), the mass of the 
ejecta $M$$_{\rm shell}$ can be determined. This is found to have a value 
$M$${_{\rm shell}}$ = 1.1$\times$10${^{-6}}$ $M$${_{\odot}}$. 
Similar calculations, as described above, have been computed for 
different temperatures yielding a  shell mass ranging between 
$M$${_{\rm shell}}$ = 1.67$\times$10${^{-5}}$ to  
1.7$\times$10${^{-7}}$ $M$${_{\odot}}$ 
for a variation in the temperature between $T$ = 2600 to 5000 K respectively.
Within the uncertainties of the parameters used, a reasonable constraint for 
the shell  mass would be in the range between 10${^{-7}}$ to
10${^{-5}}$ $M$${_{\odot}}$. The derived value of $M$${_{\rm shell}}$ 
compares  reasonably well with lower values of envelope masses determined 
for novae. In novae an average value of  
$M$${_{\rm shell}}$ = 10${^{-5}}$ $M$${_{\odot}}$ 
can be taken though deviations  by more than an order 
of magnitude  (on both lower and higher side) are often seen (Williams 1994).
It may be noted that $M$$_{\rm shell}$ derived here may be underestimated
because we have assumed  that the TiI emission originates entirely from the 
matter ejected in the present eruption and not from any pre-existing material.
Further, even if the  TiI emission arises only from the ejected material, it 
could be arising only from a  fraction of it. Thus the shell mass 
may be underestimated and this may be borne in mind while trying to classify
V838 Mon into known categories of eruptive variables - an aspect which
is discussed later in Sect. 4. 
\subsection{CO Bands}

 The other prominent feature of the $K$ band spectra are the $^{12}$CO bands.
 As shown in Fig. 3, the first overtone $^{12}$CO  ($\Delta$$v$ $=$ 2) 
 bands are seen very strongly in  V838 Mon along with the weaker $^{13}$CO
 ($\Delta$$v$ $=$ 2) bands. The presence of these CO
 bands was  first seen on 12 January ( Geballe et al. 2002a).
   The $^{12}$CO  bands   show a  complex evolution with time. 
 In the 2 February spectra, recorded at around the time of the massive
   second outburst, the 2-0    band has a deep drop at the 
2.293 ${\rm{\mu}}$m bandhead but then does    not rise back to the
 continuum level. Instead it levels off till the 3-1 band head is encountered. 
 We have verified this in both the spectra we have of this  epoch.
 Although the reason for such a behavior is not too clear to us, a tentative 
 explanation for this leveling could be that
 the individual rotational lines of the R branch which makeup the band 
 are deeply saturated.  Another unusual aspect is that the 25 March, 9 April
 and 2 May   spectra show a raised emission plateau between the
 bands - a fact first    noticed in the 9 March observations of Geballe
 et al. (2002b). This implies     that there is an emission component in 
 the CO also. Such emission is generally seen in post-AGB stars and is 
 attributable to mass loss (Oudmaijer et al.,  1995).
  High resolution  (R = 50,000) FTS spectra of
  13 February by Hinkle et al. (2002) show the individual rotational 
 lines of the $^{12}$CO  2-0 band to have a nominal heliocentric velocity 
    of -10 km/s. Such a small radial velocity would imply that the
   deep absorption component of the $^{12}$CO bands are likely to be 
 photospheric. The additional emission component must therefore 
 arise from the circumstellar ejecta. It is therefore seen that
 the CO band structure in V838 Mon is multicomponent and rather complex
 and needs to be modeled carefully. Such a model can try to
 determine factors such as the excitation temperature of the different
 CO components, the isotopic ratio of $^{13}$C/$^{12}$C in V838 Mon ( the
 solar value is 1:90 ) and also the kinematics of the ejected matter.

\subsection{Near Infrared photometry }
Finally, we use our photometry results to discuss the evolution of
 V838 Mon as seen from the shape of its near-IR continuum. The details of
the photometry are given in Table 2. 

\begin{table}[h]
\caption[]{$JHK$ photometry of V838 Mon}
\begin{tabular}{llll}
\hline\\
Obs. date (UT)&  $J$ & $H$  & $K$ \\
\hline 
\hline \\ 
 3.614 May 2002& 5.15 $\pm$ 0.14   & 4.16 $\pm$ 0.15   & 3.63 $\pm$ 0.14  \\
14.604 May 2002& 5.43 $\pm$ 0.04   & 4.43 $\pm$ 0.07  & 3.72 $\pm$  0.03 \\

\hline
\end{tabular} 
\end{table}

The $E(B-V)$  value for
 V838 Mon is  uncertain and lies between 0.25 to 0.8 ( Munari et al. 2002a).
 Following Munari et al. (2002a), we have adopted a midpoint value  of 0.5.
 We have used Koornneef's (1983) relations viz. 
  $A$${_{\rm V}}$ $=$ 3.1$E(B-V)$, 
 $A$${_{\rm J}}$ $=$ 0.265 $A$${_{\rm V}}$,
  $A$${_{\rm H}}$ $=$ 0.155 $A$${_{\rm V}}$ and
 $A$${_{\rm K}}$ $=$ 0.090 $A$${_{\rm V}}$ to correct for interstellar
 extinction. Absolute flux calibration was done by adopting zero magnitude
  fluxes from Koornneef (1983). The broad band $JHK$ fluxes from V838 Mon
   are shown in Fig. 5
  as filled circles. These broad band fluxes were fitted by  black body curves.
Different black body curves generated for temperature  increments of
100 K were tried - the best fit was decided by least squares
minimization of the deviation. Black body fits of 2600 and 2400 K are seen
to fit the data reasonably well for the 3 and 14 May data respectively.
A temperature of 2600 K for 3 May is identical with the findings  
of Munari et al. (2002b) 
at around the same time. The overall temporal evolution of V838 Mon since 
outburst has been towards cooler temperatures. This is confirmed by a large
body of photometric data as given by Munari et al. (2002a, b) and 
also from reports 
in several IAU circulars. We have  superposed the $JHK$ spectra of 2 and 
14 May in Fig. 5 so as to juxtapose them side by side. This enables to 
highlight the strong dips seen in the spectra of 2 and 14 May between 
the $J$ \& $H$ and the $H$ \& $K$ bands. However in doing so, we assume 
that errors caused by applying photometric data of 3 May to the spectra of 
2 May are marginal because of the small time-difference involved. 
The strong absorption bands between the near-IR bands is attributed to water
vapor in V838 Mon  which if present in the atmospheres of cool stars is known
to cause deep and broad absorption features at 1.4 ${\rm{\mu}}$m and
 1.9 ${\rm{\mu}}$m ( Lancon  \& Rocca-Volmerange 1992; Terndrup et al. 1991). 
  These absorption  features are seen
to be most prominent in giant stars  even cooler than M5 (Lancon \&
Roca-Volmerange 1992). However, an additional
factor that could enhance the $H$ band hump is  the H${^{\rm -}}$ ion 
which has an opacity minimum at 1.6 ${\rm{\mu}}$m. It must be mentioned that
a slightly elevated $H$ band continuum near 1.6 ${\rm{\mu}}$m was also seen
during the 26 January observations of Lynch et al. (2002).  

\begin{figure}[h]
\centering
\includegraphics[bb=137 229 399 567,width=3.1in,clip]{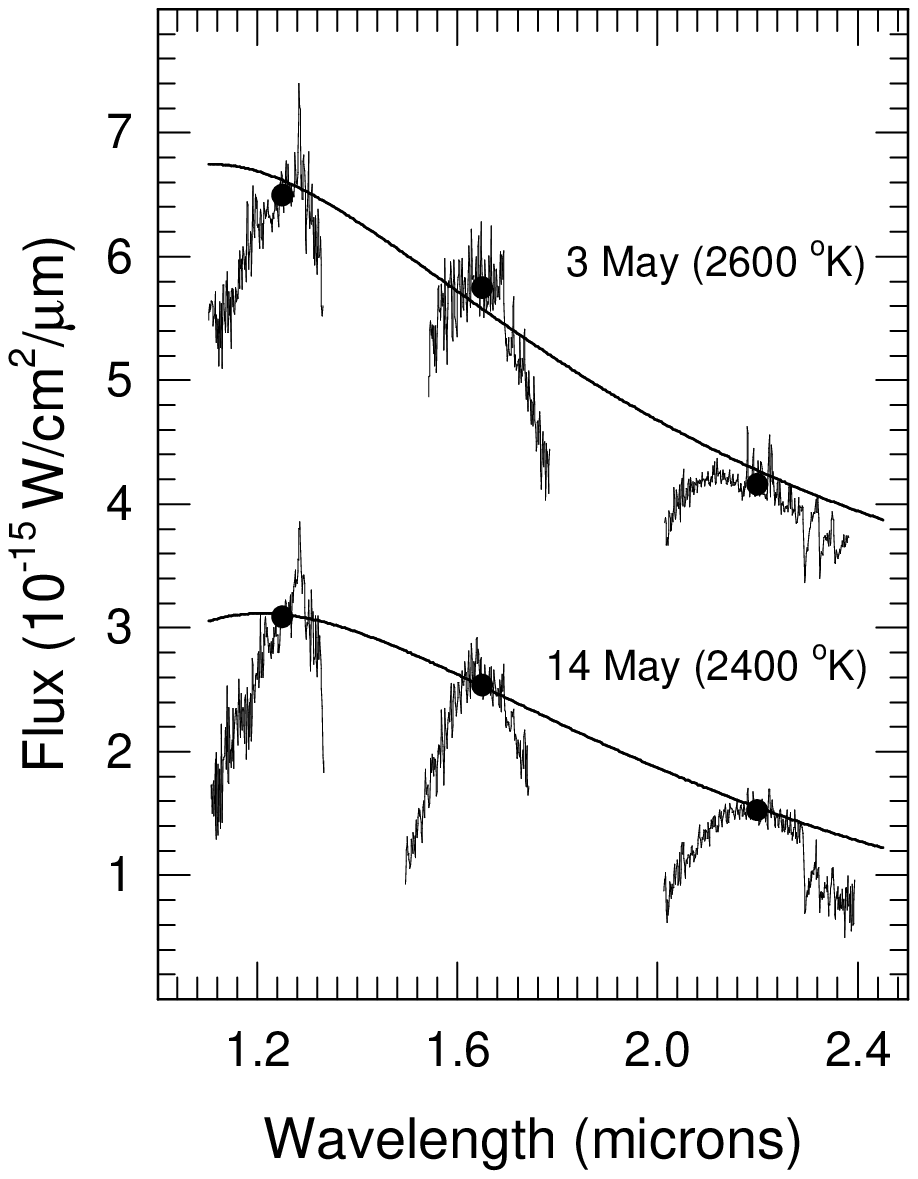}
\caption[]{ The computed fluxes from V838 Mon on 3 and 14 May 2002 as
derived from broad-band $JHK$ photometry are shown here by the filled circles.
The flux for the 14 May data can be read off directly from the graph whereas
the 2 May data has been off-set by 2.5 flux units for
clarity. Blackbody fits to these fluxes, at temperatures of 2600 and 2200 K,
are shown by the smooth bold lines. The $JHK$ spectra, represented by the 
irregular wavy lines, are  also superposed in the figure. See the text
for additional details.}
\label{fig5}
\end{figure}

\section{Discussion}       
 From the near-IR data it is thus seen that V838 Mon, near 130 days after 
outburst, shows an energy distribution similar to a very cool M-type giant.
However, many of the spectral features seen here like the TiI lines 
and the structure of the CO bands are markedly different. In many 
of the IAU circulars issued on V838 Mon, the exact nature of the 
object has been debated. Munari et al. (2002a) have dubbed the 
object's outburst
as `mysterious' and have summarized the difficulties in classifying 
it within the existing categories of eruptive variables. Apart from the 
similarities and differences brought out by Munari et al. (2002a, b),
 we see what light our
 near IR data can shed on this issue. The $JHK$ spectra of V838 Mon are
  certainly different from those of classical novae. First the Hydrogen 
  Brackett series lines are not seen and  Paschen beta is much weaker
 in strength than expected.
OI emission lines in the near-IR ( for example the 1.1287
 and 1.3166 ${\rm{\mu}}$m lines), another common feature of
 novae spectra, are also not present. 
Further, we are not aware of CO bands being seen in absorption in 
novae - in emission they are seen but rarely (for e.g. Evans et al. 1996).
Classical novae also evolve towards higher stages of ionization with time 
( the nebular and coronal phases) but there is  no indication of such an
evolutionary trend here. 

V838 Mon does not also fit easily into the scenario of a born again
AGB star like FG Sge, V605 Aql and Sakurai's object. Most of the discrepancies
that arise in such a scheme are given by Munari et al. (2002a) including 
the fact that the rise to maximum is 
too fast in comparison with known cases. To this we add that not only 
the rise time, but also the decline time of V838 Mon is  much faster than
the known cases. For the rise and decline timescales, the reader may refer
to Duerbeck et al. (2000) for Sakurai's object, Harrison (1996) for V605 Aql
 and Fig. 1 of  Blocker \& Schonberner (1997) for FG Sge. 

 Bond et al. (2002) have proposed that the closest resemblance of V838 Mon 
 is to a  red variable star (M31 RV) that erupted in M31 (Rich et al. 1989;  
 Mould et al. 1990). This view
  is shared  by Munari et al. (2002a) who also cite the outburst of V4332 Sgr
 (Martini et al. 1999) as being similar to the present case and propose that
  V838 Mon, M31 RV   and V4332 Sgr belong to a new genre of astronomical
 objects. Unfortunately there is no near-IR spectroscopic data for
 either M31 RV or V4332 Sgr with which we can  compare our spectra to
 check for similarities. However, there is a point of dissimilarity between
 M31 RV and V838 Mon that arises from our data. As mentioned earlier, in
  the case of V838 Mon an envelope mass of $\sim$ 10$^{-7}$ to 
  10${^{-5}}$ $M$${_{\odot}}$ was found - a value comparable to
   novae shell masses. In the the case of M31 RV the mass of the ejected shell
 is estimated to be much higher i.e in the range of 
0.1 to 0.001 $M$${_{\odot}}$.
  These figures are arrived at by assuming 
  equipartition between the kinetic energy 
of the shell and the total radiative energy i.e by equating the total energy 
of 10$^{46}$ ergs radiated in the first 100 days
 (Mould et al. 1990)  to the kinetic energy of
the shell  computed for expansion velocities in the range 100 - 500 km/s.
Iben \& Tutukov (1992) have modeled M31 RV in the scenario of
 a very cold white dwarf accreting
matter at a very slow rate from its binary companion and show that under
such conditions thermonuclear runaway will not take place until the accreted 
mass is much  larger than in models which represent the novae phenomenon.
As a result  larger envelope masses and outburst luminosities  are predicted
 - as is found in the case of M31 RV. How V838 Mon fits into such a model
needs to be addressed and worked out. Other similarities and differences
between V838 Mon, M31 RV and V4332 Sgr have been brought out by
 Munari et al (2002a) and  Martini et al. (1999). Based on such comparisons,
 there is a fair  possibility that all  three objects belong to a new class of
 astronomical objects.

  \begin{acknowledgements}
	  
	  The research work at Physical Research Laboratory is funded by
the Department of Space, Government of India. We thank Varricatt W.P. of
Joint Astronomy Centre, Hawaii for help in photometric reductions. We
thank the anonymous referee whose constructive comments helped improve 
the paper. This work has made use of
 data available from the ADC and CDS data centers and also from data 
 available at the following websites viz. \protect http://www.noao.edu/archives.html, \protect http://kurucz.harvard.edu/linelists.html and 
\protect  http://www. kusastro.kyoto-u.ac.jp.vsnet.
\end{acknowledgements}


\end{document}